\documentclass[aps,prl,twocolumn,superscriptaddress,showpacs]{revtex4}

\usepackage{avant}
\usepackage[T1]{fontenc}

\usepackage{amsmath,amssymb,graphicx}
\usepackage{color,textcomp}
\usepackage[squaren]{SIunits}
\usepackage{subfigure}
\usepackage[normalem]{ulem}

\newcommand {\Frechet}{Fr\'{e}chet }

\begin{document}

\title{
Molecular self-organization: predicting the pattern diversity and lowest energy state of competing ordering motifs 
}
\author{B. A. Hermann}
\email{b.hermann@cens.de}
\affiliation{Center for Nano Science (CeNS) and Walther-Mei{\ss}ner-Institute of Low Temperature
Research of the Bavarian Academy of Sciences,
Walther-Meissner-Str. 8, 85748 Garching, Germany}
\author{C. Rohr}
\affiliation{Center for Nano Science (CeNS) and Walther-Mei{\ss}ner-Institute of Low Temperature
Research of the Bavarian Academy of Sciences,
Walther-Meissner-Str. 8, 85748 Garching, Germany}
\author{M. Balb\'{a}s Gambra}
\affiliation{Arnold Sommerfeld Center for Theoretical Physics (ASC) and Center for Nano Science (CeNS), Department of Physics,
 LMU M{\"u}nchen, Theresienstra{\ss}e 37, 80333 M{\"u}nchen, Germany}
\author{K. Gruber}
\affiliation{Center for Nano Science (CeNS) and Walther-Mei{\ss}ner-Institute of Low Temperature
Research of the Bavarian Academy of Sciences,
Walther-Meissner-Str. 8, 85748 Garching, Germany}
\author{A. Malecki}
\affiliation{Center for Nano Science (CeNS) and Walther-Mei{\ss}ner-Institute of Low Temperature
Research of the Bavarian Academy of Sciences,
Walther-Meissner-Str. 8, 85748 Garching, Germany}
\author{M.~S.~Malarek}
\affiliation{Department of Chemistry, University of Basel, Spitalstrasse 51,
4056 Basel, Switzerland}
\author{E. Frey}
\affiliation{Arnold Sommerfeld Center for Theoretical Physics (ASC) and Center for Nano Science (CeNS), Department of Physics,
 LMU M{\"u}nchen, Theresienstra{\ss}e 37, 80333 M{\"u}nchen, Germany}
\author{T. Franosch}
\affiliation{Institut f\"ur Theoretische Physik, Universit\"at Erlangen-N\"urnberg, Staudtstra{\ss}e 7, 91058 Erlangen, Germany}
\affiliation{Arnold Sommerfeld Center for Theoretical Physics (ASC) and Center for Nano Science (CeNS), Department of Physics,
 LMU M{\"u}nchen, Theresienstra{\ss}e 37, 80333 M{\"u}nchen, Germany}

\date{\today}

\begin{abstract}
Self-organized monolayers of highly flexible \Frechet dendrons were deposited on graphite surfaces by solution casting. Scanning tunneling microscopy (STM) reveals an unprecedented variety of patterns with up to seven stable hierarchical ordering motifs serving as a versatile model system. The essential molecular properties determined by molecular mechanics simulations are condensed to a coarse grained interaction site model of various chain configurations. In a Monte Carlo approach with random starting configurations the experimental pattern diversity can be reproduced in all facets of the local and global ordering. Based on an energy analysis of the Monte Carlo and molecular mechanics modeling the thermodynamically most stable pattern is predicted coinciding with the pattern, which dominates in the STM images after several hours or upon moderate heating.

\end{abstract}





%
%

\maketitle

\begin{figure}
     	\includegraphics[width = 0.85 \columnwidth]{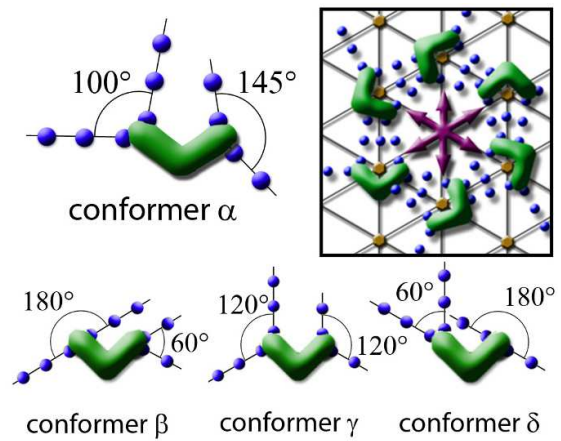}
       	\caption{
       	(color online) Top: angle configuration for conformer $\alpha$. The box displays the six possible orientations on a fixed lattice of hexagonal symmetry. Bottom: Conformers $\beta$, $\gamma$, and $\delta$ with the respective internal angles indicated.
       	}
      	\label{fig:conformers}
\end{figure}

\begin{figure}[width = columnwidth]
     \includegraphics [width = 0.85 \columnwidth] {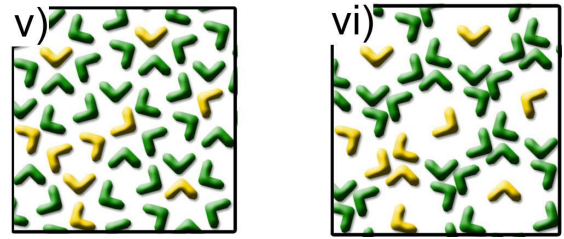}
     \caption{
(color online) Alternative phases predicted condensed from random starting configurations by the Monte Carlo approach:, but not found in the experiment: v) honeycomb pattern of conformer $\alpha$ and $\gamma$, and vi) inverted honeycomb pattern of conformer $\gamma$, can be condensed from the random starting configuration. Both the inverted honeycomb and honeycomb MC patterns proved unstable upon simulating a finite temperature and thus cannot be considered as likely patterns; these patterns find no close resemblance in the experimental data.
     }
     \label{fig:MC_modeling_unstable}
\end{figure}

\begin{figure}[width = columnwidth]
     \includegraphics [width = 1.0 \columnwidth] {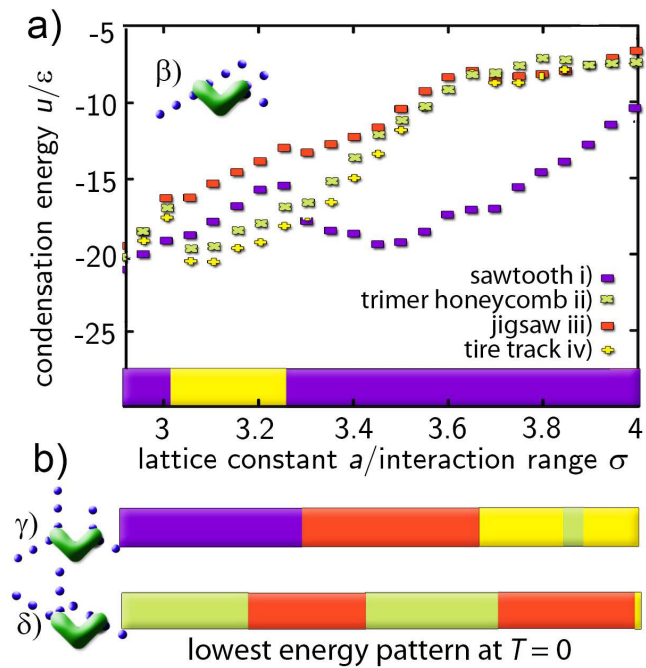}
     \caption{
(color online) conformer $\beta$, coarse grained model with angles 180$^\circ$ and 60$^\circ$ between the long and short chains; a.) zero temperature energy analysis for determining the associated lowest energy state; inset: triangular lattice displaying the six possible orientations . The experimental relevant regime corresponds to $3.2  a/\sigma$. b.) The pattern of lowest energy in the zero temperature analysis for conformations $\gamma$ and $\delta$. At the experimentally relevant regime of $3.2  a/\sigma$, the sawtooth pattern for conformer  $\gamma$ and jigsaw pattern for conformer $\delta$, did not prove stable upon heating and hence cannot represent the thermodynamically most stable pattern.
     }
     \label{fig:zero_temp}
\end{figure}

Organic molecules mainly self-organize via hydrogen bonds,~\cite{Pawin:2006,Schnadt:2008} metal-coordination~\cite{Langner:2007} and van der Waals interactions~\cite{Tahara:2006}. Voices have been raised,~\cite{Barth:2007,Tomba:2009} that more innovative modeling approaches~\cite{Shi:2006,Ilan:2008} are necessary in order to analyze and predict molecular patterns for fostering applications in surface functionalization~\cite{Barth:2005}, sensors~\cite{Lehn:2002}, catalysis~\cite{Lorenzo:2000}, and in molecular electronic devices~\cite{Joachim:2000}. Predictability could speed up molecular design and advance the understanding of the self-organization process itself. For \emph{predicting} several coexisting ordering motifs (cf. ~\cite{Morgenstern:2003,Meier:2005}) a new multi-modeling approach is required, that does not rely on the prior knowledge of the resulting patterns~\cite{Rohr:2010,Weber:2008, Haran:2007,Sarlah:2005}. Here we show, that based on a coarse grained interaction site model (omitting chemical details) we are able to not only reproduce and predict all features of the local/global ordering motifs of self-organized molecular layers, but also independently predict the thermodynamically most stable pattern and thus deliver an innovative approach to the understanding of molecular self-organization.\\      
\begin{figure*}
 \begin {minipage} [tc] {0.30 \hsize}
    \includegraphics[width = 5.4 cm]{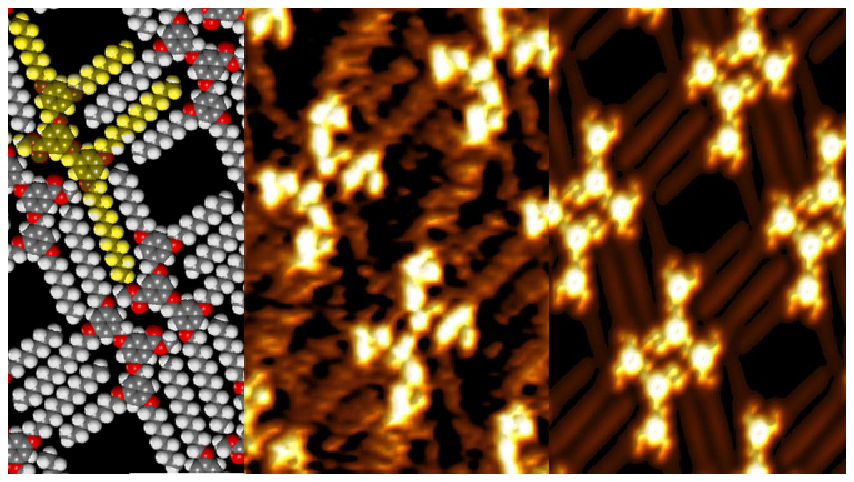}
       \caption{
       (color online) The dodecyl/octyl terminated \Frechet dendron is highlighted in yellow. One of seven different ordering patterns on HOPG is exemplified: \emph {middle}: STM image of a jigsaw pattern of the \Frechet dendron  (12.5~nm $\times$ 7~nm, $U_{Bias}$ = -800~mV, $|I_T|$ = 8~pA). \emph {left}: Atomistic modeling by an energy minimized MM simulation (see text). \emph {right}: simulation of the LDOS by DFT calculations based on the input of the geometry determined by MM.
       }
      \label{fig:STM_MM_DFT}
 \end {minipage}
 \begin {minipage} {0.8 \hsize}
 \end {minipage}
 \begin {minipage} [tl] {0.60 \hsize}
    \includegraphics [width = 11.4 cm] {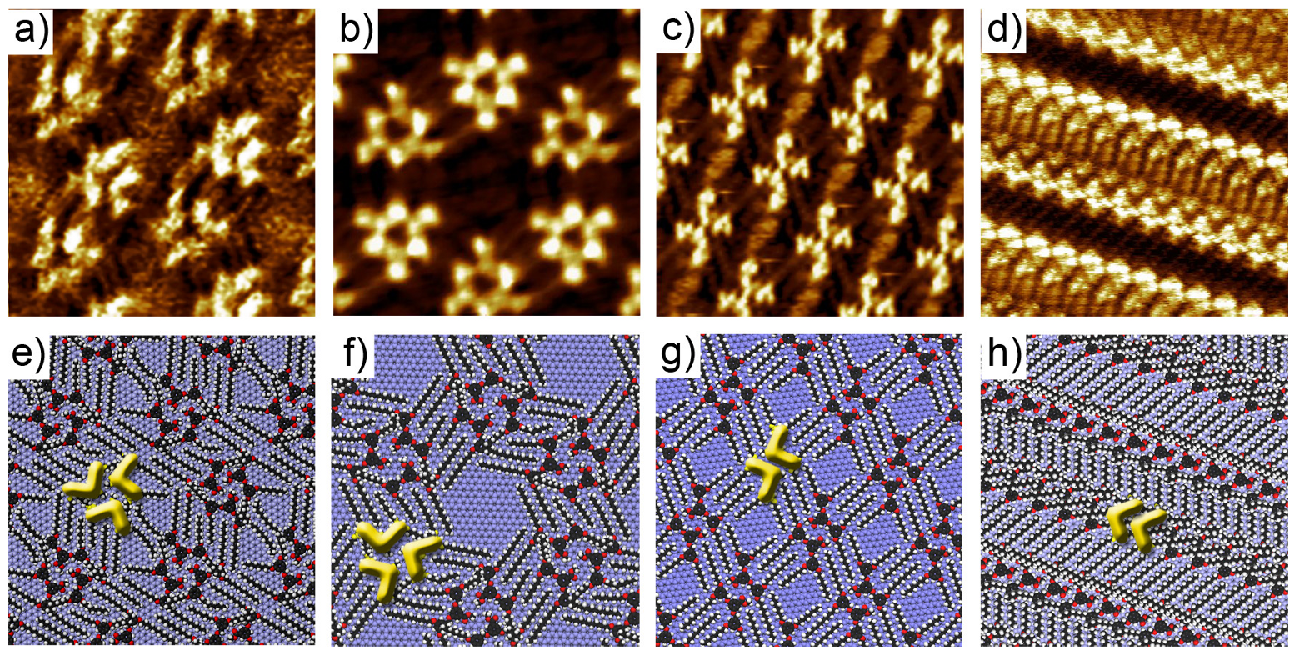}
     \caption{
(color online) (a-d) STM results and corresponding (e-f) molecular mechanics modeling (see text)  of four experimentally found patterns of \Frechet dendrons : a) sawtooth b) trimer honeycomb, c) jigsaw, d) tiretrack. The molecular backbones marked in yellow indicate the local ordering motif. (12.5~nm $\times$ 22~nm, $U_{Bias}$ = -700 to -800~mV, $|I_T|$ = 5 to 30~pA) 
     }
     \label{fig:MM_modeling}
 \end{minipage}
\end{figure*}
In this article, the self-organization of flexible supra-molecular building blocks, \Frechet dendrons~\cite{Hawker:1990,Hermann:2006}, is investigated in highly resolved scanning tunneling micrographs and modeled with various techniques: density functional theory (DFT), molecular mechanics (MM) and Monte Carlo (MC)  simulation based on an interaction site model. The second generation asymmetric \Frechet dendrons consist of three phenyl rings, two of which are elongated by two pairs of alkoxy chains, a longer 12mer pair and a shorter 8mer pair (see Fig.~\ref{fig:STM_MM_DFT} left), responsible for van der Waals and weak hydrogen bonds ~\footnote{ omitted due to a much smaller contribution}. All van der Waals interactions present are the sum of the intermolecular interaction (chain - neighboring - chain) and the molecule-substrate interaction (molecule - substrate). By varying the internal angles of the pairs of alkoxy chains in an interaction site model, all facets of the experimentally found patterns can be predicted within our approach. A zero temperature analysis of the MC simulations points to a thermodynamically most stable pattern in line with conclusions of the local energetics derived from MM simulations and experimental findings upon heating. \\
For the STM measurements a Nanoscope Multimode III equipped with a low-current converter under ambient conditions is employed. Monolayers of \Frechet dendrons (3,5-bis[(3,5-bisoctyloxyphenyl)methyloxy]phenyl esther, synthesis see ~\cite{Constable:2008}), are cast from 0.2 mM solutions in ethanol, hexane or heptadecane on highly oriented pyrolytic graphite (HOPG) surfaces. The emerging patterns are analyzed by combining several complementary approaches: a.) DFT, utilizing the CASTEP $\textregistered$ module of Material Studio 4.3 employing Perdew-Wang '91 generalized gradient approximation exchange correlation functionals~\cite{Clark:2005} and a plane wave basis set with an energy cutoff at 260~eV; b.) MM, performed on a fixed double-layer of graphite (universal force fields~\cite{Rappe:1992} in periodic boundary conditions) based on starting values obtained from the STM experiments; c.) MC annealing simulations in Metropolis scheme based on 60$^\circ$ rotations of a coarse grained interaction site model, using periodic boundary conditions for minimizing finite-size effects in system sizes ranging from several hundreds to a few thousand objects. \\
The mixed graph of Fig.~\ref{fig:STM_MM_DFT}~\footnote{Images flattened, correlation averaged, drift-corrected.} exemplifies MM and DFT simulations together with measurements of the jigsaw pattern. After applying a droplet of the molecules in hexane, the jigsaw pattern (\emph{p2} symmetry)~\cite{Constable:2007} appears as one of seven general ordering motifs in about 25\% of the substrate area. A detailed analysis of many STM images yields starting configurations~\footnote{starting with an energy-minimized free molecule has failed at 150\% minimization energies} for a MM energy minimization~\footnote{ unit cell of $2$ molecules in periodic boundary conditions.} to calculate molecular conformations (see left of Fig.~\ref{fig:STM_MM_DFT}) and relative positions. From the MM determined adsorption energy the dominant van der Waals part~\footnote{electrostatic interactions contribute only 30\%.} can be extracted, which will be discussed later.\\
As the energy-minimized MM simulations can only provide atomic positions and not electronic states, we used these energy minimized geometries as input to derive the local density of states (LDOS) of a free single molecule using DFT. Convoluted~\footnote {with a Gaussian function of 2$\times$ a Pt-$d_z$-orbital, 200~pm} density contours of a planar slice in the LDOS are plotted~\footnote {$8$ free molecules at precisely MM determined positions} (right, Fig.~\ref{fig:STM_MM_DFT}) for a direct comparison with the STM measurement: the phenyl rings of the molecular cores are discernible as three bright protrusions and atoms of the alkoxy arms emerge (visible in the dark areas) as series of faint lines (right, Fig.~\ref{fig:STM_MM_DFT}) in agreement with the experimental STM images (middle, Fig.~\ref{fig:STM_MM_DFT}).\\
In a large STM study, we have imaged seven ordering motifs of this \Frechet dendron in excellent resolution. Here we focus on four main ordering motifs: sawtooth : trimer  honeycomb : jigsaw : tiretrack (see Fig.~\ref{fig:MM_modeling}). 20~nm $\times$ 20~nm sized domains of these patterns in ratios 1\%~:~15\%~:~25\%~:~20\% typically cover an HOPG surface half an hour after applying a hexane droplet. About 39\% of the HOPG surface is occupied by domain boundaries, mobile molecules or two other patterns: upright and small tiretrack. Minutes after casting a droplet of the \Frechet dendron in hexane, the honeycomb pattern dominates. The sawtooth pattern seldom appears as an independent pattern and often as a stacking fault of the honeycomb pattern. After several hours typically 60\% of the HOPG surface is covered with the tiretrack pattern in hexane. \\
As DFT calculations can only poorly describe van der Waals interactions, the dominant part of the internal energy, the adsorption energy per unit cell or area, is evaluated from the MM minimization energy~\footnote{omitting electrostatic interactions}. This covers the dominant intermolecular interactions stemming from the alkoxy chains, statistically contributing an interaction strength of 14 kJ/mol per four CH$_2$-units for more than half but not fully interdigitated alkoxy chains on graphite (previous studies yield 18 kJ/mol for fully and 8.5 kJ/mol for half-interdigitated alkoxy chains on graphite surfaces~\cite{Yin:2001}). We have calculated the van der Waals part of the total energy I) of a substrate supported monolayer in MM energy minimized geometry, $i_{sup}$,  II) of a gas phase net~\cite{Schnadt:2008}, $i_{gn}$, and III) of one isolated molecule, $i_{iso}$. With that we can carefully separate the intermolecular interaction (chain - neighboring-chain) part $I_{c-n-c}/$molecule$ = (i_{gn}/n)- i_{iso}$ from the substrate interaction (molecule - substrate) part $I_{m-s}/$molecule$ = (i_{gn}- i_{sup})/n$ of the van der Waals energy per molecule, with $n$, the number of molecules per unit cell (see Table I a). With dividing accordingly by the area per unit cell, the respective energies per nm$^{2}$ are derived (see Table I b).\\
Molecular domains grow or dissolve by outer molecules, which have only a fraction $\delta <1$ of the total number of neighboring molecules, adsorbing or going over into a mobile state, respectively. When initially forming domains after applying a droplet, the gain in van der Waals adsorption energy $I_{m-s}+\delta (I_{c-n-c})$ \emph{per molecule} rules the process: the honeycomb pattern in table~I~a. Over time the HOPG surface is covered with some 100~nm$^2$ sized domains, which transform according to the Kitaigorodskii principle of avoiding free space~\cite{Kitaigrorodskii:1973}. Hence the highest gain in van der Waals adsorption energy $I_{m-s}+I_{c-n-c}$ \emph{per area} dominates the outcome: the tiretrack pattern in table~I~b. This agrees with the thermodynamically stable pattern experimentally identified by moderate heating, which is the tiretrack pattern in hexane. Thus, the initial domains form open pore patterns dominated by the highest energy gain \emph{per molecule}, while with time the pattern with the largest energy gain \emph{per area} wins. However, all calculated energies are very close, reflecting that the phases can coexist.\\
  \begin {table}
  	\caption{
Van der Waals part of the adsorption energies determined by MM energy minimization. a) \emph{per molecule}. b) \emph{per area}. The labels refer to: m-s, molecule-substrate interactions and c-n-c, chain-neighboring-chain interactions.
}
 		\begin{ruledtabular}
			\begin{tabular}{cccc}
a)						& molecules	& $I_{m-s}$		& $I_{c-n-c}$	\\
            	& per unit	& /molecule		& /molecule		\\
pattern      	&		cell		& [kJ/mol]		& [kJ/mol] \\
\hline
		sawtooth 	&		6				& -674				& -147			\\
		honeycomb &		6				& -691				& -163			\\
		jigsaw 		& 	2 			& -653				& -117			\\
		tiretrack	& 	2 			& -528				& -222			\\

\hline
\hline
b)						& unit cell & $I_{m-s}$ 	& $I_{c-n-c}$ \\
        			&  size     &  /nm$^2$ 		& /nm$^2$ 		\\
pattern				&  [nm$^2$] & [kJ/mol] 		& [kJ/mol] \\
\hline
		sawtooth 	& 	24.5 		& 	-167			& 	-38			\\
		honeycomb & 	26.5 		& 	-155			& 	-38			\\
		jigsaw 		& 	8.4 		& 	-151			& 	-25			\\
		tiretrack & 	6.1 		& 	-172			& 	-75			\\

			\end{tabular}
		\end{ruledtabular}
 \end{table}
In order to identify the ground state energy we performed independent MC simulations, based on coarse graining the dominant molecule and substrate properties in an interaction site model. Rather than including as much chemical detail as possible, we investigated the crucial properties reproducing key experimental findings. When large molecules adsorb on ten or more substrate atoms the organization is dominated by steric hindrance; the role of the substrate is to align the molecular backbone and to some extend the molecular chains. In this work, we systematically varied the internal angles of the pairs of longer and shorter arms of interaction centers (see Fig.~\ref{fig:conformers}), thereby probing different molecular conformations, in order to determine how critical the angles are on the predictive power of the interaction site model. \\
As in our previous work~\cite{Rohr:2010} the substrate symmetry is accounted for by allowing discrete $\pi/3$ rotations of the coarse grained molecules, see Fig.~\ref{fig:conformers} a, on fixed sites in a lattice of hexagonal symmetry. The geometry and spacing of the molecular backbone is transferred from the MM modeling. The chains are modeled by a small number of beads, disposed in rigid, straight arms at distances determined by MM and in ~\cite{ Rohr:2010} approximate angles of 100$^\circ$ and 145$^\circ$ between long and short pairs of alkoxy chains, respectively. We refer to a \Frechet-dendron with the latter chain conformation as conformer \textbf{$\alpha$} in this article. Here, the internal angle within each pair of arms has been systematically varied from 180$^\circ$ and 60$^\circ$, referred to as conformer $\beta$, over 120$^\circ$ and 120$^\circ$, conformer $\gamma$, to 60$^\circ$ and 180$^\circ$, conformer $\delta$, for long and short pairs of alkoxy chains, respectively. One bead represents four CH$_2$-units, exhibiting short-range van-der-Waals attractions, which are described by a Lennard-Jones potential, $U(r) = 4 \epsilon[ (\sigma/r)^{12}-(\sigma/r)^6]$. \\
\begin{figure}[width = columnwidth]
     	\includegraphics[width = 0.85 \columnwidth]{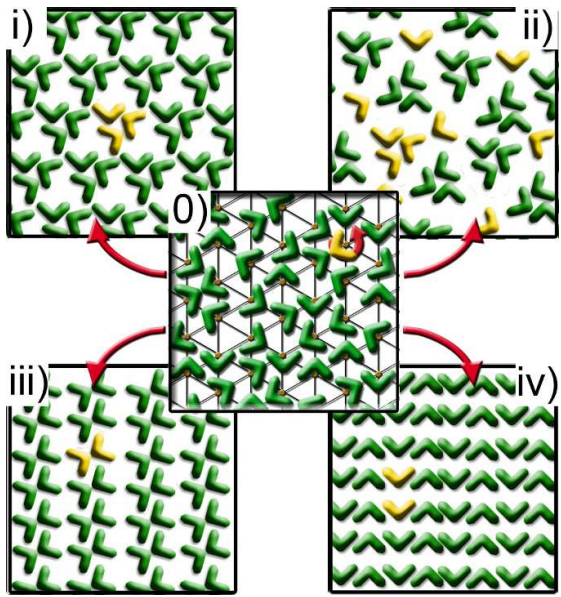}
       	\caption{
       	(color online) Monte Carlo (MC) simulations (i-iv). 0) Random starting configurations allowing discrete $\pi/3$ rotations of the coarse grained molecules. Upon slow cooling the following patterns formed from random state (0) indicated with red arrows: i) Saw tooth pattern for conformer $\beta$ and for conformer $\alpha$, ii) trimer honeycomb pattern for conformer $\delta$, iii) jigsaw pattern for conformer $\alpha$, iv) tiretrack pattern for conformers $\alpha$, $\beta$, and $\gamma$. All phases were stable upon heating to 300$^\circ$C.
       	}
      	\label{fig:MC_modeling}
\end{figure}
The MC simulations were performed by preparing random starting configurations (see o) in Fig.~\ref{fig:MC_modeling}) and relaxing the system by slowly "cooling". The lattice constant $a$, accounting for the density, has been varied between $2.95 \,\sigma$ and $4.0\,\sigma$, corresponding to a change in coverage density by a factor of 2.5, while the dimensionless inverse temperature $\epsilon/k_B T$ ranged from $0.025$ to $6$. The ordered patterns obtained: i), ii), iii), iv) in Fig.~\ref{fig:MC_modeling}, correspond well in long-range symmetry and local ordering with the structures found experimentally a), b), c), d) in Fig.~\ref{fig:MM_modeling}, respectively. While the chain configuration of conformer $\alpha$ (100$^\circ$ and 145$^\circ$ for long and short pairs of alkoxy chains, respectively) leads to a striking variety of patterns (see~\cite{ Rohr:2010}), the systematic choice of internal angles in conformers $\beta$, $\delta$ and $\gamma$ narrows the pattern variety. \\
The patterns that can be generated from the respective conformers are: sawtooth and tiretrack for $\beta$ (see i) and iv) in Fig.~\ref{fig:MC_modeling}), tiretrack, honeycomb and inverted honeycomb for $\delta$ (see iv) in Fig.~\ref{fig:MC_modeling} and v) and vi) in Fig.~\ref{fig:MC_modeling_unstable}, respectively) and trimer honeycomb for $\gamma$ (see ii) in Fig.~\ref{fig:MC_modeling}). So with the latter conformer $\gamma$ even the intriguing open-pore host structure of molecular trimers arranged in hexagonal symmetry (see b) in Fig.~\ref{fig:MM_modeling}) can now be generated from random starting configurations by slow cooling. The trimer honeycomb pattern b) in Fig.~\ref{fig:MM_modeling} displays an organizational chirality with an included guest molecule of non-fixed orientation (described in another experiment~\cite{Merz:2005}). Hence all experimentally found patterns can be completely described.  The MC patterns i) to iv), in very good correspondence with experimental findings, proved to be reliable, as raising the temperature showed that these phases stay ordered up to 300$^\circ$C and can hence be considered as stable states. \\
Additionally, we exemplify here two more patterns (see Fig.~\ref{fig:MC_modeling_unstable}) named v) honeycomb pattern and vi) inverted honeycomb pattern (both found for conformer $\gamma$), that were condensed from random starting configurations, but could not be verified in the STM experiments. The stability of these MC patterns was tested by simulating a finite temperature; both patterns proved unstable upon "heating" and thus cannot be considered likely ordering states. Though, all three honeycomb patterns have a similar appearance, upon closer examination, only the trimer honeycomb pattern ii) displayed in Fig.~\ref{fig:MC_modeling} correctly describes the experimentally found pattern b) of Fig.~\ref{fig:MM_modeling}. Hence, the honeycomb and inverted honeycomb pattern find no explicit resemblance in the experimental data and are unstable in a MC simulation of finite temperature. \\
In contrast to patterns generated from random starting configurations, patterns i), ii), iii), iv) (see Fig.~\ref{fig:MC_modeling}) can be pre-prepared for all conformers allowing to compute the associated energies in a zero temperature energy analysis. So the state of lowest energy at $T~=~0$ can be identified in each case as function of the density. In the case of artificially prepared patterns the energy analysis can only indicate how energetically attractive a structure is, not if the pattern is likely. The pattern that minimizes the energy varies, depending on the packing fraction (see bar graphs in Fig.~\ref{fig:zero_temp} a) and b) and the internal angle, hence, the conformer. In the experimentally relevant regime ($ 3.2 a/\sigma$), the most favorable configuration for conformer $\alpha$ (see~\cite{ Rohr:2010}) and conformer $\beta$) is the tiretrack pattern (see Fig.~\ref{fig:zero_temp} a). As this pre-prepared pattern is stable upon "heating" to 300$^\circ$C and in a second way can be generated from random starting configurations, the tiretrack pattern is identified as the thermodynamically most stable pattern in nice agreement with the experimental findings and the MM energy analysis discussed above. \\
For the chain configurations of conformer $\delta$ and $\gamma$ the sawtooth and trimer honeycomb patterns (Fig.~\ref{fig:zero_temp} b)) appear to have the lowest energy at $T~=~0$ in the experimentally relevant regime, respectively. Probing the stability of the latter two patterns by raising the temperature leads to a destabilization and thus clearly indicates that neither the sawtooth nor trimer honeycomb pattern can represent the thermodynamically most stable pattern around room temperature. \\
In summary, the systematic variation of internal angles of the pairs of chains (representing the chain-chain interdigitation) reveals that the local and global ordering motif of all experimentally found patterns can be reproduced by cooling random starting configurations. The zero temperature analysis of pre-prepared patterns in all combinations of conformers/patterns underlines that only the tiretrack pattern found for conformer $\alpha$ (see~\cite{Rohr:2010}) and $\beta$ represents an ordered state with lowest energy at $T~=~0$ and is at the same time stable upon "heating". Thus, the tiretrack pattern denotes experimentally and theoretically the thermodynamically most stable motif at finite temperatures, even for different molecular conformations in the interaction site model. \\
Yet, patterns cannot only be reproduced by the MC approach, but also predicted. In Fig.~\ref{fig:MC_modeling} iv) a second local ordering motif with a  side by side arrangement of the molecular backbones is apparent (Fig.~\ref{fig:wave_pattern} a). This local ordering motif named "wave" has been discovered after the theoretical prediction; it occurs rarely for dodecyl/octyl terminated \Frechet dendrons in ethanol (Fig.~\ref{fig:wave_pattern}). \\

Our study underlines that main geometrical features are the dominant driving force of molecular self-organization. The various patterns found experimentally for the \Frechet dendron presented in this study serve as a versatile model system to systematically test our multi-modeling approach. Five structural phases have been closely modeled by MM energy minimization; their quality is underlined exemplarily by a DFT LDOS simulation plotted adjacent to the STM image. Evaluating the van der Waals part of the MM determined adsorption energy favors an open pore trimer honeycomb structure in the initial phase of adsorption (energy per molecule maximized) and a densely packed pattern reminiscent of tire tracks after various phase transformations in the final state (energy per area maximized). The calculated energies are very close reflecting the coexistence of phases for competing ordering motifs.  Moreover, we have refined our independent MC approach with a coarse grained interaction site model of various internal angles and so successfully reproduce the experimental pattern diversity in all facets of the local and global ordering. The lowest energy pattern at $T=0$ can be identified in a zero temperature analysis for the different chain conformations. All theoretical and experimental findings point to the tiretrack pattern as thermodynamically most stable pattern for the dodecyl/octyl terminated \Frechet dendron. Furthermore, a newly predicted local ordering motif has been experimentally verified afterwards. We thoroughly and successfully tested an interaction site approach to molecular self-organization, which will serve as an innovative analysis tool in the future fostering the application of self-organized molecular monolayers in various areas of science. \\

\begin{figure}[width = columnwidth]
     \includegraphics [width = 0.9 \columnwidth] {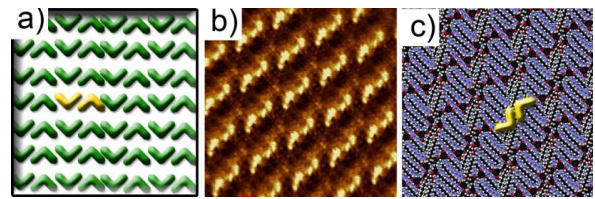}
     \caption{
(color online) a.) Another local ordering motif of the MC simulation Fig.~\ref{fig:MC_modeling} iv.)   found afterwards by b.) STM: measurement of a rarely found fifth pattern with corresponding b.) MM modeling.( 12.5~nm $\times$ 22~nm, $U_{Bias}$ = -800~mV, $|I_T|$ = 8~pA)
     }
     \label{fig:wave_pattern}
\end{figure}

\begin{acknowledgments}
Financial support by the German Excellence Initiative via the program ``Nanosystems Initiative Munich (NIM)'', Studienstiftung des dt.\ Volkes, IDK NanoBioTechnology, ERA-Chemistry, CeNS, NRP47  and the Swiss Nanoscience Institute  is gratefully acknowledged.
\end{acknowledgments}



\end{document}